\documentclass[twocolumn,preprintnumbers,amsmath,amssymb,prl,superscriptaddress]{revtex4-1}

\usepackage{amssymb}
\usepackage{amsmath}
\usepackage{graphicx}
\usepackage{bm}
\usepackage{natbib}
\usepackage{physics}

\begin{document}
\title{Pressure-resistant intermediate valence in Kondo insulator SmB$_6$}

\author{Nicholas P. Butch}
\email{nicholas.butch@nist.gov}
\affiliation{Center for Nanophysics and Advanced Materials, Department of Physics, University of Maryland, College Park, MD 20742}
\affiliation{NIST Center for Neutron Research, National Institute of Standards and Technology, 100 Bureau Drive, Gaithersburg, MD 20899}
\affiliation{Lawrence Livermore National Laboratory, 7000 East Ave., Livermore, CA 94550}
\author{Johnpierre Paglione}
\affiliation{Center for Nanophysics and Advanced Materials, Department of Physics, University of Maryland, College Park, MD 20742}
\author{Paul Chow}
\author{Yuming Xiao}
\affiliation{HP-CAT, Geophysical Laboratory, Carnegie Institute of Washington, Argonne, Illinois, 60439, USA}
\author{Chris A. Marianetti}
\affiliation{Department of Applied Physics and Applied Mathematics, Columbia University, New York, NY 10027, USA}
\author{Corwin H. Booth}
\affiliation{Chemical Sciences Division, Lawrence Berkeley National Laboratory, Berkeley, CA 94720}
\author{Jason R. Jeffries}
\affiliation{Lawrence Livermore National Laboratory, 7000 East Ave., Livermore, CA 94550}
\date{\today}

\begin{abstract}
Resonant x-ray emission spectroscopy (RXES) was used to determine the pressure dependence of the f-electron occupancy in the Kondo insulator SmB$_6$. Applied pressure reduces the f-occupancy, but surprisingly, the material maintains a significant divalent character up to a pressure of at least 35~GPa. Thus, the closure of the resistive activation energy gap and onset of magnetic order are not driven by stabilization of an integer valent state. Over the entire pressure range, the material maintains a remarkably stable intermediate valence that can in principle support a nontrivial band structure.

\end{abstract}

\maketitle

The study of intermediate valent compounds, which dates back half a century \cite{Menth69}, has recently been reinvigorated by the possibility that Kondo insulators harbor a topological surface state \cite{Dzero10}. The best-studied candidate material is SmB$_6$ \cite{Alexandrov13,Lu13}, which was only recently shown to be electrically insulating in the bulk \cite{Zhang13,Kim13,Wolgast13}. Although activated electrical transport is readily inferred from the dramatic temperature-dependent resistivity, at low temperatures the resistivity saturates at a finite value instead of diverging to infinity. This anomalous behavior is now linked to the presence of electrically conducting surface states \cite{Syers15}, and many experiments have been directed at determining the topological classification of the material \cite{Neupane13,Frantzeskakis13,Jiang13,Zhu13,Li14,Biswas14,Xu14,Phelan14,Ruan14,Fuhrman15,Nakajima16,Wolgast15}. A full description of the underlying strongly correlated electron state is key to understanding nontrivial topology in f-electron materials.

Applied pressure is a powerful tool for studying intermediate valent systems because their electronic states are sensitive to small changes in interatomic separation. A well-known example of this behavior is SmS, in which the divalent ``black'' phase is dramatically transformed into a semiconducting intermediate valent ``gold'' phase by the application of very modest pressure of 0.6~GPa \cite{Jayaraman70}. As pressure increases, Sm becomes fully trivalent and it magnetically orders \cite{Barla04,Deen05,Annese06}. SmB$_6$ follows this example: an intermediate valence exists already at ambient pressure \cite{Chazalviel76}, and the valence is sensitive to temperature \cite{Mizumaki09} and chemical pressure \cite{Tarascon80}. Applied pressure stabilizes a metallic and magnetic ground state over the range 4-10~GPa \cite{Cooley95,Barla05,Derr08} and increases the Sm valence  \cite{Rohler87,Beaurepaire90}. By analogy, a pressure-induced trivalent state is anticipated in SmB$_6$, but it has not yet been shown spectroscopically, nor is it known how its onset correlates with either metallization or magnetic order \cite{Barla05}.

Unexpectedly, our experiment shows that an intermediate valent state in SmB$_6$ persists to the highest measured pressures, about 35~GPa, and shows no signs of tending towards saturation. This unprecedented discovery implies that neither the metallization nor the onset of magnetic order are associated with simple integer valency, but are characteristics of a robust intermediate-valent state. Such a state violates the paradigm that valence fluctuations destabilize spins on long time scales, and demands a new model of f-electron valence stability. Most intriguingly, the coexistence of magnetism and topologically nontrivial intermediate bulk valence \cite{Alexandrov13} may help explain recent suggestions of metallic magnetism on the surface of SmB$_6$ \cite{Nakajima16,Wolgast15}.

\begin{figure}
\begin{center}
\includegraphics[width=3.3in]{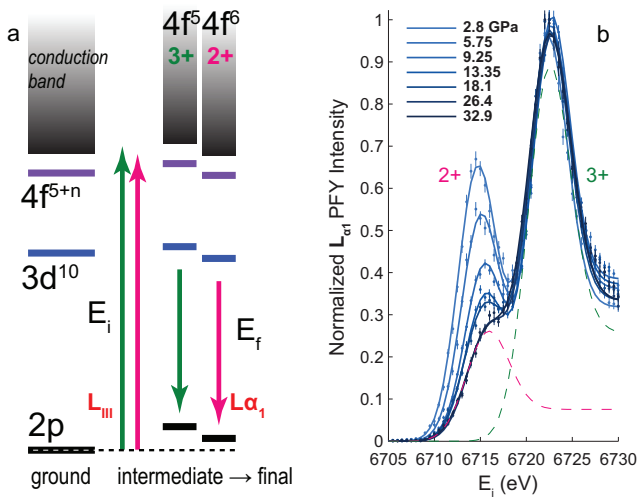}
\end{center}
\caption{a) Schematic of Sm core-hole excitations probed by the PFY and RXES measurements, including incident and emitted x-rays $E_i$ and $E_f$. b) Partial fluorescence yield data at different pressures, normalized to the maximum value at the 3+ peak. The peak attributed to the divalent configuration clearly decreases with increasing pressure, but does not vanish. Individual valence contributions to the peak-edge structure 32.9~GPa are indicated by dashed lines and correspond to Eqn.~\ref{eqPFY}. Error bars denote an uncertainty of one standard deviation.}
\label{PFY}
\end{figure}

X-ray emission spectroscopy, which probes the local electronic configuration of atoms, is a valuable discriminator of multivalent ions. The measurement involves two coupled transitions: 1) an electron in a low-lying core orbital is excited into the conduction band via absorption of an x-ray photon, and 2) an electron from another orbital decays to fill the hole, emitting a lower-energy x-ray photon. The final state of the system is excited with respect to the initial state. A schematic of the excitations studied in our experiment is presented in Fig.~\ref{PFY}a. The removal of the core electron from the 2p shell, corresponding to the L$_\mathrm{III}$ absorption edge, requires an incident energy $E_i \approx 6720$~eV, the exact value of which depends on the 4f shell electron occupancy. In intermediate-valent Sm compounds, x-ray absorption spectroscopy (XAS) measurements yield two peaks, split by 7 eV, that correspond to integer-occupancy f$^5$ and f$^6$ configurations \cite{Mizumaki09}. This is due to the short-lived, local nature of the core-hole excitation and its f-occupancy-dependent Coulomb screening. One possible relaxation pathway is for an electron to decay from the 3d$_{5/2}$ to 2p$_{3/2}$ shell, emitting a photon with energy $E_f = 5636$~eV (the L$\alpha_{1}$ emission line) that is independent of the 4f shell occupancy.

We performed X-ray Diffraction (XRD), Partial Fluorescence Yield (PFY), and Resonant X-ray Emission Spectroscopy (RXES) measurements under pressure at the HPCAT beamline at the Advanced Photon Source. The XRD measurements were performed on a powder of crushed single crystals of SmB$_6$ \cite{Zhang13} using a 29.2~keV x-ray beam. For the inelastic measurements, a flux-grown 50~$\mu$m single crystal was studied using a 35~$\mu$m diameter x-ray beam in a diamond anvil cell sealed with a Be gasket. Approximately hydrostatic pressure conditions were achieved using neon as a medium, and manometry via ruby fluorescence indicated typically 5\% uncertainty in the pressure.

The PFY experiment consists of a measurement of the intensity at fixed $E_f$ as a function of scanned $E_i$. The intensity is proportional to the absorption probability, but the measured emission linewidths have the advantage of being intrinsically sharper compared to standard XAS measurements (compare to \cite{Mizumaki09}). Because the x-ray absorption process represents an ejection of a photoelectron, details of the absorption edge reflect the unoccupied density of states. In SmB$_6$, a large unoccupied density of states due to a d-derived band leads to a peak structure similar to that seen in other f-electron compounds. The pressure dependence of L$\alpha_{1}$-PFY scans across the L$_\mathrm{III}$ edge are presented in Fig.~\ref{PFY}b. The presence of two prominent peaks corresponding to different integer f-occupancy is consistent with prior x-ray absorption and photoelectron spectroscopy on SmB$_6$ \cite{Chazalviel76}.

Quantitatively, the PFY intensity $I_\mathrm{PFY}(E_i) = \sum a_v N_v$ includes a contribution from each valence configuration $v$ well-described by the sum of Gaussian and sigmoid functions centered at energy $E_v$. For $\varepsilon_v = \frac{1}{\sqrt{2}W}(E_i - E_v)$,
\begin{equation} \label{eqPFY}
N_v =  \left( \frac{e^{-\varepsilon_v^2}}{ \sqrt{2\pi} W} \right)_\mathrm{peak} + r \left( \frac{1}{2} + \int_0^{\varepsilon_v} d\zeta \frac{e^{-\zeta^2}}{2\sqrt{\pi}} \right)_\mathrm{step}
\end{equation}
where $2 \sqrt{2 \ln{2}} W$ is the full width at half maximum, and $r$ is the ratio of the Gaussian and sigmoid amplitudes. Example fits to the total intensity, as well as the contribution of each valence, are shown in Fig.~\ref{PFY}b. The weighted ratio of the amplitudes $(2 a_2 + 3 a_3)/(a_2+a_3)$ yields the value of the intermediate valence. As applied pressure increases, $a_2$ decreases relative to $a_3$, reflecting an increasing valence.

\begin{figure}
\begin{center}
\includegraphics[width=3.3in]{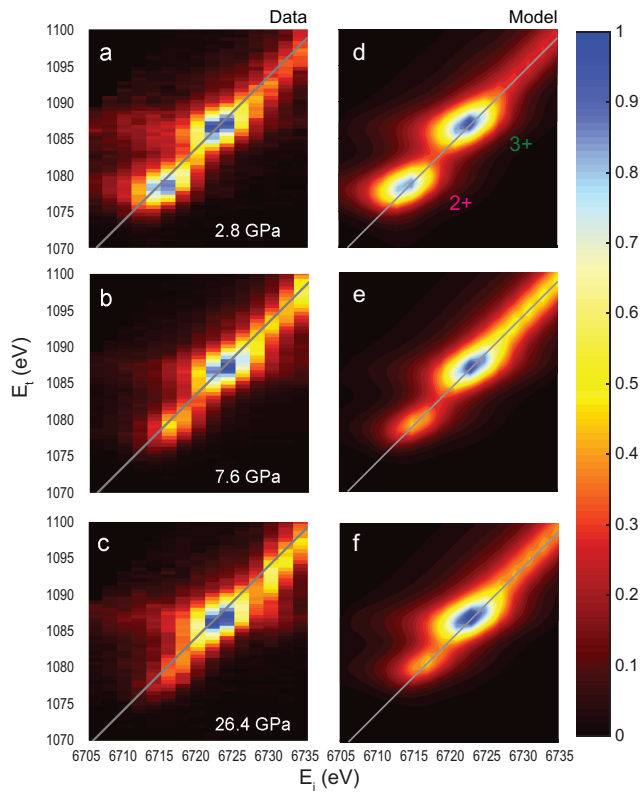}
\end{center}
\caption{Comparison of RXES data (a-c) and fits (d-f) at different pressures. With increasing pressure, the lower energy peak associated with the divalent state is suppressed. Also clearly evident are the resonances at constant $E_t$. The gray lines denote constant values of $E_f$ corresponding to the fluorescence line. Data are fit to Eqn.~\ref{eqRXES} as described in the text. Intensities are normalized to the maximum intensity at the 3+ peak.}
\label{RXES}
\end{figure}

To get a complete picture of the energy dependence of the resonant absorption-emission process, we also performed RXES measurements, in which both $E_i$ and $E_f$ are varied. Data from RXES scans at different pressures are shown in Fig.~\ref{RXES} as a function of $E_i$ and the transferred energy $E_t = E_i - E_f$. For reference, a gray diagonal line having constant $E_f$ indicates the trajectory of the previously-discussed PFY measurements. The RXES lineshape resembles the PFY, broadened along $E_i$ and $E_f$.

The RXES intensity $I_\mathrm{2D}(E_i,E_t)$ is fit using the Kramers-Heisenberg formula for photon-atom scattering \cite{Kvashnina11,Booth14}:
\begin{equation} \label{eqRXES}
\begin{aligned}
I_\mathrm{2D} = \int d\epsilon N(\epsilon) \Phi \frac{1 }{\pi \frac{\Gamma_{i}}{2} (1 + x^2)} \frac{1}{\pi \frac{\Gamma_{t}}{2} (1 + y^2)}
\end{aligned}
\end{equation}
where $x = \frac{2}{\Gamma_{i}} (E_{i} - E_{0} - \epsilon)$ and $y = \frac{2}{\Gamma_{t}} (E_{t} - E_{0} - \epsilon)$ for transitions with energy $E_0$. The unoccupied density of states $N(\epsilon)$ is convolved with two-dimensional Lorentzian broadening having full widths at half maximum $\Gamma_{i}$ and $\Gamma_{t}$. $N(\epsilon)$ follows the form of Eqn.~\ref{eqPFY}, and the total intensity is a sum of two terms with spectral weight proportional to the relative contribution of each valence configuration. The amplitude $\Phi \propto |\bra{f}T_2 \ket{i} \bra{i}T_1 \ket{g}|^2$ is determined by the transition probabilities between ground, intermediate, and final states. Its value is approximately constant over the relevant energies, although an asymmetry along $E_i$ is well described by assigning to $\Phi$ a skewness term $1 + \frac{2}{\pi} \atan(\lambda x)$, with $\lambda \approx -0.3$ \cite{SI}.

\begin{figure}
\begin{center}
\includegraphics[width=3.3in]{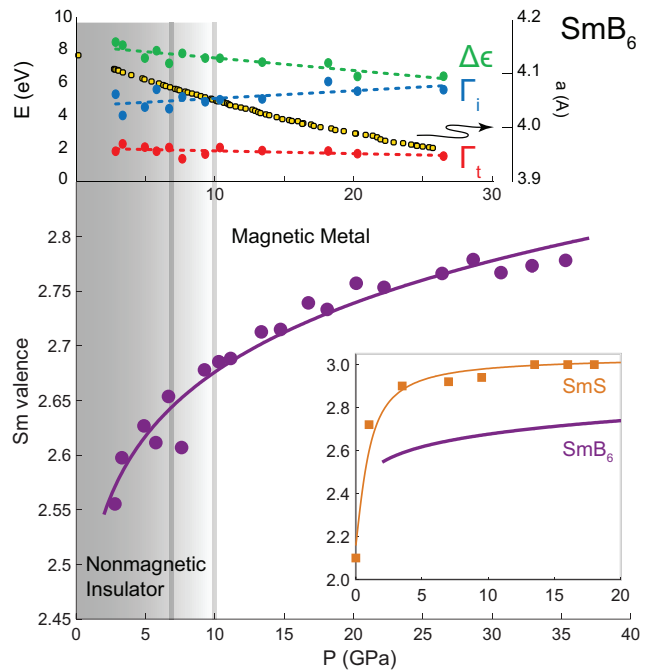}
\end{center}
\caption{Pressure dependence of the room-temperature Sm valence in SmB$_6$. The smooth pressure dependence is insensitive to the transition from an insulating nonmagnetic ground state to magnetic metal, which occurs over a pressure range indicated by the gray lines, and a robust intermediate valent state is maintained. Top: The energy parameters derived from fits to Eqn.~\ref{eqRXES} and the lattice parameter also exhibit a smooth pressure dependence independent of the change in ground state. Inset: This behavior contrasts starkly with the archetypal pressure-induced valence transition in SmS \cite{Annese06}. Uncertainty of one standard deviation in the valence is smaller than the plotted points. }
\label{phsdgm}
\end{figure}

The contribution from the 2+ peak in the RXES spectrum clearly diminishes with increasing pressure.  The amplitudes from the fits to Eqn.~\ref{eqRXES} support the intermediate valence derived from our PFY analysis, and such an interpretation of RXES data is also previously established \cite{Kvashnina11,Hayashi13}. Figure~\ref{phsdgm} summarizes our findings, including the pressure dependence of several important energy scales. The valence at 10~GPa is 0.1 smaller than determined by earlier XAS measurements \cite{Rohler87,Beaurepaire90}. The slope of the valence as a function of pressure is greatest near 2.5, a behavior attributable to near-degeneracy of the divalent and trivalent configurations \cite{Zell81}. The splitting $\Delta \epsilon$ between absorption edges for each valence decreases from 8.5~eV to 6.5~eV between ambient pressure and 26 GPa, reflecting less of a difference in screening between the two configurations. The width $\Gamma_i$ increases from 5~eV to 6~eV, while the value for $\Gamma_t$ remains constant at 2~eV. These linewidths are consistent with those derived from studies of other f-electron compounds \cite{Booth14}.

These findings challenge conventional understanding of intermediate valent insulators. The monotonic pressure dependence of the valence in SmB$_6$ is insensitive to both the closing of the hybridization gap and sharp onset of long-range magnetic order, and other energy scales are also continuous through these transitions. This is remarkable because both the Fermi surface and magnetic ground state are expected to be determined by f-electron fluctuations, as seen in other rare earth compounds. Most striking is that the Sm valence in SmB$_6$ not only never reaches 3+, but remains strongly intermediate valent at high pressures, defying predictions \cite{Barla05,Ogita05,Derr08,Nishiyama13}. The unusual electronic configuration is surprisingly robust.

Fluctuations of the charge configuration of unpaired f-electrons affect the spin state, reducing the paramagnetic response and impeding magnetic order \cite{Riseborough80}. A well-known example is Ce metal, which undergoes an isostructural volume collapse under pressure that changes its valence and suppresses the Curie-Weiss paramagnetic response \cite{Dallera04}. Both metallic and insulating Kondo lattice compounds can be described by a periodic Anderson model, a fact exploited in the classification of topological Kondo insulators \cite{Dzero10}. As a practical matter, although there are many examples of metallic intermediate valent compounds, whose designation overlaps with heavy fermion materials, very few intermediate valent compounds exhibit electrical insulating behavior down to low temperatures \cite{Riseborough00}.

The archetypal intermediate valent insulator is SmS, in which pressure-induced magnetic order and metallization are associated with the onset of full 3+ valence. This material exhibits a prominent first order transition in the lattice constant at 1~GPa, which is coupled to a large jump in valence from 2 to 2.7 \cite{Deen05,Annese06}. Above 2~GPa, magnetic order sets in \cite{Barla04} inside the intermediate valent regime, but the magnetic phase does not occupy the entire volume until 5 GPa, when the sample is nearly trivalent. A coincidence between volume change, metallization, and valence change is observed also in Sm, Tm, and Yb monochalcogenides \cite{Jarrige13,Tsiok14}.

The trend in pressure-tuned ground state is obeyed by SmB$_6$, as it becomes magnetically ordered below 12~K \cite{Barla05,Derr06} once the insulating gap closes by 10~GPa \cite{Derr08}. Yet despite the similar phase diagrams of SmB$_6$ and SmS, the pressure dependence of the Sm valence is very different (Fig.~\ref{phsdgm} inset), and our experiment shows that the inconsistency with integer valence in SmB$_6$ under pressure is more substantial than suggested previously \cite{Barla05,Ogita05}. The root of this discrepancy lies in the pressure dependence of the SmB$_6$ lattice constant determined via XRD (Fig.~\ref{phsdgm}), which decreases smoothly as a function of applied pressure \cite{Nishiyama13} to values greater than 25~GPa \cite{Parisiades15} and does not collapse. Indeed, the lattice of SmB$_6$ may be considered already collapsed, as the bulk modulus \cite{Parisiades15} is three times larger than that of SmS \cite{Hailing84}.

A correspondence between lattice parameter and valence is an established component of the intermediate valence phenomenology. It intuitively derives from the different radii of the stable integer valent ions, which contract with increasing valence. In SmB$_6$, this rule has been inferred from comparisons of XAS and magnetometry on chemically substituted samples that demonstrate a relation between lattice constant and valence \cite{Tarascon80,Beaurepaire90}. It also appears to be responsible for the unusual ambient-pressure negative coefficient of thermal expansion \cite{Mandrus94} that is accompanied by a valence decrease as the temperature is lowered \cite{Mizumaki09}. Yet, these arguments do not simply extend to the pressure data; already by 10 GPa the experimentally determined lattice constant is 4.05~{\AA} (Fig.~\ref{phsdgm}), far smaller than the 4.115~{\AA} value of hypothetical trivalent SmB$_6$ derived from substitution studies \cite{Tarascon80}. The failure of ion size alone to determine valence is reminiscent of the limitations of simple promotional models to describe gradual valence changes beyond the Sm monochalcogenides \cite{Jarrige13}, at which point it becomes necessary to invoke hybridization to describe the intermediate valent state. The fact that SmB$_6$ is already far from integer valence at ambient pressure underscores that correlations play an essential role in determining the electronic state \cite{Lu13,Kang15}. Our results highlight the need for new theoretical insight into why the SmB$_6$ valence is sensitive to temperature but less so to applied pressure, and what underlying interactions are responsible for the metallization and onset of magnetic order.

Finally, we address efforts to experimentally determine the topological classification of SmB$_6$. Because theoretical calculations suggest that SmB$_6$ is topologically nontrivial at all experimentally relevant values of valence \cite{Alexandrov13}, pressure is not expected to tune the material through a topological transition until integer valence is achieved. In principle, opening an energy gap at the Fermi level at any pressure converts SmB$_6$ into a topological insulator, and if the onset of magnetic order could be decoupled from metallization, a strain-engineered interface between magnet and topological insulator could be used to stabilize exotic edge states with potential use in future devices. In this light, we suggest that recent observations of one-dimensional surface transport \cite{Nakajima16} and unusual magnetotransport \cite{Wolgast15} could be consistent with the existence of a strain-stabilized magnetic surface on SmB$_6$. Our results demonstrate that a surface having strongly intermediate valence, as has been detected spectroscopically \cite{Heming14}, can support magnetic order.

\begin{acknowledgments}
NPB acknowledges support by CNAM and the LLNL PLS directorate. JRJ is partially supported by the Science Campaign. Portions of this work were performed under LDRD (Tracking Code 14-ERD-041). This work was performed at HPCAT (Sector 16), Advanced Photon Source (APS), Argonne National Laboratory. HPCAT operations are supported by DOE-NNSA under Award No. DE-NA0001974 and DOE-BES under Award No. DE-FG02-99ER45775, with partial instrumentation funding by NSF. The Advanced Photon Source is a U.S. Department of Energy (DOE) Office of Science User Facility operated for the DOE Office of Science by Argonne National Laboratory under Contract No. DE-AC02-06CH11357. LLNL is operated by Lawrence Livermore National Security, LLC, for the DOE, NNSA under Contract No. DE-AC52-07NA27344. Crystal growth at UMD was supported by AFOSR-MURI (FA9550-09-1-0603). CAM was supported by the NSF MRSEC program through Columbia in the Center for Precision Assembly of Superstratic and Superatomic Solids (DMR-1420634). Work at Lawrence Berkeley National Laboratory was supported by the Director, Office of Science (OS), Office of Basic Energy Sciences (OBES), Chemical Sciences, Geosciences, and Biosciences Division of the U.S. Department of Energy (DOE) under Contract No. DE-AC02-05CH11231.
\end{acknowledgments}

\bibliography{SmB6bib}

\end{document}